\documentclass[12pt,preprint]{aastex}

\newcommand{\obja}{QSO PG1718+4807}
\newcommand{\object}{QSO PG1718+4807}
\newcommand{\zabs}{$z_{\rm abs}$}

\newcommand{\lya}{Ly$\alpha$}

\newcommand{\lyaf} {\lya\ forest}

\newcommand{\kms}{km s$^{-1}$}

\newcommand{\lnhi}{$\log N_{HI}$}

\newcommand{\cmm}{cm$^{-2}$}

\begin{document}
\title{New HST spectra indicate the \obja\ will not give the primordial
deuterium abundance\altaffilmark{1}}

\author{ David Kirkman\altaffilmark{2,3,4}, David
 Tytler\altaffilmark{2,3}, John M. O'Meara\altaffilmark{2,3}, \\ Scott
 Burles\altaffilmark{2,5}, Dan Lubin\altaffilmark{2,3}, Nao
 Suzuki\altaffilmark{3,3}, Robert F. Carswell\altaffilmark{6}, Michael
 S. Turner\altaffilmark{7}, \& E. Joseph Wampler\altaffilmark{8}
\altaffiltext{1}{Based on observations obtained with the NASA/ESA
Hubble Space Telescope obtained by the Space Telescope Science
Institute, which is operated by AURA, Inc., under NASA contract
NAS5-26555.}
\altaffiltext{2} {Visiting Astronomer, W.M. Keck Observatory which
is a joint facility of the University of California, the California
Institute of Technology and NASA.}
\altaffiltext{3} {Center for Astrophysics and Space Sciences,
University of California, San Diego,
MS 0424; La Jolla; CA 92093-0424}
\altaffiltext{4} {E-mail: david@mamacass.ucsd.edu}
\altaffiltext{5} {Experimental Astrophysics, Fermi 
National Accelerator Laboratory, Box 500, Batavia, IL 60510-0500}
\altaffiltext{6} {Institute of Astronomy, University of Cambridge,
Madingley Road, Cambridge, CB3 0HA}
\altaffiltext{7} {Department of Astronomy and Astrophysics,
Enrico Fermi Institute, 5640 S. Ellis Ave., The University of
Chicago, Chicago, IL 60637-1433}
\altaffiltext{8} {2386 Empire Grade Rd., Santa Cruz, CA 95060-9701}
}

\begin{abstract}

The \zabs $\sim$ 0.701 absorption system towards \obja\ is the only
example of a QSO absorption system which might have a
deuterium/hydrogen ratio approximately ten times the value found
towards other QSOs.  We have obtained new STIS spectra from the Hubble
Space Telescope (HST) of the \lya\ and Lyman limit regions of the
system.  These spectra give the redshift and velocity dispersion of
the neutral hydrogen which produces most of the observed absorption.
The \lya\ line is too narrow to account for all of the observed
absorption.  It was previously known that extra absorption is needed
on the blue side of the main H~I near to the expected position of
deuterium.  The current data suggests with a 98\% confidence level
that the extra absorption is not deuterium.  Some uncertainty persists
because we have a low signal to noise ratio and the extra absorption
-- be it deuterium or hydrogen -- is heavily blended with the \lya\
absorption from the main hydrogen absorption.

\end{abstract}


\section{Introduction}

Only a few percent of QSOs have an absorption system which shows
deuterium.  Ground based spectra of most QSOs at redshift $z \sim 3$
show at least one absorption system having enough H~I that the D~I
lines could be detected in spectra with resolution $\sim 10$ \kms\ and
signal to noise ratio (SNR) $\sim 50$. Such absorption systems always
have enough H~I to show Lyman continuum absorption and are called
Lyman limit systems (LLS).  Unfortunately, most LLS do not show
deuterium because there is too much H~I absorption near the expected
deuterium position: --82 \kms\ towards shorter wavelengths
from the H~I.  To date, only five systems at high redshift have been
shown to either measure or to limit the primordial D/H abundance, and
all agree with a value of D/H = $3.0 \pm 0.4
\times 10^{-5}$ (\citet{dodorico01}, \citet{omeara01},
\citet{kirkman00}, \citet{burles98a}, \citet{burles98b}).

Several groups have tried to find deuterium in absorption systems at
redshifts $< 2$. This has the advantage that there is much less \lyaf\
absorption at low redshift.  However, most of the H~I which hides
deuterium may be associated with the main (LLS) hydrogen absorption, and
it is not known whether such associated absorption becomes more or
less common at lower redshifts.

Only one absorption system has been found at low redshift which might
give the D/H ratio: \obja, which is the subject of this paper. The
reasons that we do not know of many more are that only about 30 QSOs
are bright enough for 10 \kms\ spectra from the HST, and LLS are rare
at low redshifts.  The LLS at \zabs $\sim$ 0.7011 towards \obja\ has
the features expected of an absorption system which could show
deuterium.  It has a very steep Lyman limit \citep{lanzetta93} and
simple metal lines (\citet{tytler99}, \citet{webb97a}), both of which
indicate that the H~I may have a simple velocity structure.  It also
has a low metal abundance: [Si/H] $\simeq -2.4$ \citep{tytler99}.

Three groups have discussed one HST GHRS spectrum which covers the
\lya\ and Si~III 1206 lines.  Webb et al. (1997a, 1997b) noted that the
\lya\ line is asymmetric and can be fit with two symmetric Voigt
profile components, separated by about 80 \kms .  If the component on
the lower wavelength side is entirely deuterium, then the entire \lya\ line
could arise from a single component, with D/H = $20 \pm 5 \times
10^{-5}$.  \citet{levshakov98} and \citet{tytler99} fit the same
spectrum with single component H+D profiles, and found a wider range
of $4.4 \times 10^{-5} <$ D/H $< 57 \times 10^{-5}$
(95\%). \citet{tytler99} also noted that all of the asymmetry could be
explained by H~I in a velocity component near -82 \kms , rather than
D~I, in which case D/H could not be measured in this absorber.

A variety of interpretations are possible because we have inadequate
spectra of \object.  First, the GHRS spectrum has low SNR: about 200
times fewer photons per \kms\ than the ground based spectra used to
measure D/H.  Second, the GHRS spectrum of \object\ included just
\lya\ and one metal line.  The D/H measurements which we have made in
absorption systems at high redshift show 16 -- 18 of the Lyman series
lines and 5 - 20 metal lines.  Both deficiencies mean that a wide
range of parameters are allowed, and that it can not be determined if
the extra absorption is deuterium or hydrogen.  In this paper we
present new HST STIS spectra of \object.

\section{Observations and data reduction}

Prior discussions of \object\ utilized both space and ground based
data.  The space based data consisted of GHRS spectra covering the
\lya\ and SiIII lines of the LLS (\citet{tytler99}, \citet{webb97a},
\citet{levshakov98}), in addition to IUE spectra which covered the
Lyman limit \citep{lanzetta93}.  The ground based data utilized the
HIRES spectrograph on the Keck-I telescope \citep{tytler99} to
cover the wavelengths encompassing the MgII absorption of the LLS.

In this paper, we present new STIS spectra from HST which cover two
regions crucial to the analysis of the D/H system towards \object:
the \lya\ line and the Lyman limit.  We also present additional
Keck HIRES spectra of the MgII absorption.

\subsection{New HST STIS spectra of \object}

We have two new sets of STIS spectra of \object: longslit FUV-MAMA
observations and echelle NUV-MAMA observations.  Here, we describe
each type of observation, and in following sections, we discuss two
instrumental issues which affect the data analysis, namely the
STIS line spread function, and wavelength discrepancies between the
STIS and GHRS spectra.

The STIS longslit exposures were taken using the FUV-MAMA detector
through the 52x0.2" slit and dispersed using the G140M grating.  The
total exposure time consisted of 11,327 seconds over four orbits on
January 12 and 13, 1999.  The spectra cover the wavelength range of
1540 -- 1594 \AA , with a pixel size of 0.0529 \AA , and a resolution
of 2.4 pixels FWHM.  The exposures were processed using the standard
CALSTIS pipeline.  They were rebinned to a common wavelength scale
then the weighted average was calculated.  In terms of the \object\
LLS, the longslit data cover the Ly-6 through Ly-18 transitions, and
wavelengths shortward of the Lyman limit.  At the Lyman limit, the
total signal to noise is approximately 8.5 per pixel.  The final
combined spectrum obtained from the long slit exposures is shown in
Figure 1.

The STIS echelle exposures were taken using the NUV-MAMA detector
through the 0.2x0.2" slit, and dispersed with the E230M echelle.  The
total exposure time consists of 14,131 seconds taken over five orbits
on November 1, 1999.  The spectra cover the wavelength range of 1871
-- 2629 \AA , with some spectral overlap between echelle orders.  The
pixel size was $\simeq 0.0336$ \AA , with a resolution of 1.9 pixels,
or a FWHM of 9.3 \kms.  The echelle data cover the \lya , SiIII, SiIV,
and C IV transitions of the LLS.  At the \lya\ transition of the LLS
near 2068 \AA , the final SNR is approximately 5 per pixel.  As with
the longslit data, the data were processed using the CALSTIS pipeline,
and were coadded in a similar fashion -- the result is shown in Figure
2.  We note that the version of CALSTIS used only a 1-D model for
scattered light removal, which may result in flux errors of up to 3 \%
near the \lya\ absorption.  While there is now a 2-D version of the
scattered light removal algorithm available from STScI, we have not
used it on our data as our signal to noise is only 5.  In both
datasets, a local continuum was fit to the data by eye, using regions
which were deemed to be free of absorption.

\subsection{Line spread functions}

When modeling the GHRS data, we used a Gaussian line spread function
(LSF) with a FWHM of 14.1 \kms.  Unlike the GHRS LSF, the STIS LSF can
not be treated as Gaussian.  Changes in the LSF cause significant
differences in the measured velocity width of absorption lines and are
of particular importance to our analyses.  Thus instead of using a
Gaussian LSF for the STIS observations, we use those provided by
\citet{sahu00} based on a combination of emission line spectra and
theoretical simulations.  The LSFs used in our analysis for the E230M
and G140M gratings are shown in Table 1.  We note that the true LSF
may be slightly asymmetric, but that this asymmetry depends upon the
orientation of the slit on the sky in a complicated way.  After
consulting with Sahu, we choose a symmetric LSF based on an averaging
of the two sides to the STIS LSF for our analysis.

\subsection{Wavelength calibration}

Between the CALSTIS pipeline reduced STIS spectra presented here and
the GHRS observations of \object\ used in Tytler et al. (1999), there
is a 0.08 \AA\, or 12 \kms\  difference in the observed wavelength of
the \lya\ transition of the LLS near 2068 \AA. This difference was
obtained via a cross-correlation of the two spectra over the common
wavelengths covered, and confirmed by fitting profiles to several narrow
absorption features present in both spectra.

The origin of this difference is unknown.  It was corrected by
shifting the GHRS spectrum to match the STIS data.  We choose the STIS
spectrum as the wavelength reference so that there would be
consistency between the redshift scale used for the high order Lyman
series lines and the \lya\ absorption.  We note that the STIS
wavelengths for the redshift of the LLS agrees with the redshift of
the MgII absorption from the HIRES data.

\section{The absorption properties of the LLS at \zabs=0.7011}

The new STIS data presented here cover the Lyman limit at high
resolution for the first time and give additional spectra of the \lya\
and metal line absorption.  We discuss first the parameters describing
the hydrogen followed by measurements and limits to the parameters
describing the metal lines of the LLS, and then the metallicity,
ionization, and thermal characteristics of the system.

\subsection{Hydrogen Absorption}

The STIS longslit data allow a good determination of the hydrogen
absorption parameters present in the LLS towards \object .  As
illustrated in Figure 1, we fit the observed Lyman series transitions
through to the Lyman limit, along with the residual flux blueward of
1554 \AA .  Also present in the spectrum is absorption due to Galactic
C IV near 1548 and 1551 \AA , which were not fit, as they did not
affect the determination of the hydrogen parameters.  All absorption
features of interest were fit using Voigt profiles convolved with the
LSF in Table 1.

From the STIS longslit spectra of the Lyman limit, we determine a
column density of \lnhi $= 17.22 \pm 0.005$ \cmm , a velocity width of
$b = 22.5 \pm 0.2 $ \kms , and a redshift of $z=0.701084 \pm
(2\times10^{-6})$.  This column density is consistent
with that determined from the IUE spectrum of the Lyman limit
(\citet{lanzetta93}, \citet{tytler99}).  The errors quoted represent
only those random errors identified in the fitting procedure and do not
include systematic errors.  One such systematic error is the placement for the
QSO continuum -- particularly over the
Lyman continuum absorption -- which may be in error by a few percent.

The portion of STIS echelle spectra that covers the \lya\ line of the
LLS is shown in Figure 2 along with the prior GHRS data for
comparison.  The new STIS spectra have higher resolution, but lower
SNR.  In Figure 3, we show two fits to the observed STIS and GHRS
spectra.  The first fit shows the absorption due to hydrogen \lya\
using the parameters determined from the Lyman limit STIS longslit
data.  This single absorber alone cannot fit all of the observed
absorption on either the red or the blue sides of the line.  The
residual blue side absorption was previously known and is of primary
interest, as it may or may not be caused entirely by deuterium
absorption.  The red side residual absorption is new result of this
work.  The results of the fits to the hydrogen are summarized in Table
2.

\subsection{Metal line absorption}

Also summarized in Table 2 are the parameters describing the metal
lines seen in the LLS.  The SiIII (1206) absorption is seen in both
the GHRS and STIS spectra, and is shown in Figure 4.  C IV is
observed in the STIS echelle data, and is found to be consistent with
the values derived from Tytler et al. (1999).  The limit on the
SiIV absorption comes from the STIS data presented here.

We re-observed the MgII absorption in \obja\ for an additional 5000
seconds with HIRES at Keck.  The data were obtained using with the
same setup and reduced in the way described in \citet{tytler99}.  The
MgII absorption from the combined data set is shown in Figure 5.  The
line parameters from the combined data set are presented in Table 2.

\subsection{The thermal state of the LLS}

As with previous studies of D/H systems, we can use the metal line
widths to estimate the thermal and turbulent velocity structure of the
gas in the LLS.  With this knowledge we hope to predict the velocity
width of the deuterium absorption and compare it with the data to
determine whether the observed absorption is consistent with
deuterium.  In ideal cases, all metal line widths are consistent with
a single thermal and turbulent solution (e.g. \citet{omeara01}.  In the
case of \object, however, we find inconsistent solutions.

There are two parts to the inconsistency: (1) The $b_{SiIII} >>
b_{MgII}$ such that both can not arise from the same gas and (2) the
$b_D$ predicted by the $b_H$ and either the $b_{SiIII}$ or the
$b_{MgII}$ is smaller than the extra absorption on the blue side of
\lya, $27.1 \pm 5.2$ \kms.  The first inconsistency leads us to
consider two models -- one using just $b_{SiIII}$ and the other using
just $b_{MgII}$.

In one model, we use the measured velocity widths of the H I and SiIII
absorption alone to determine a temperature of T$= 1.9 \times 10^{4}$
K and a turbulent velocity width $b_{turb} = 16.5$ \kms .  This
solution, shown by the upper dashed line in Figure 7, predicts $b_D =
19.2 \pm 0.84$ \kms, which is significantly smaller than the $b$ value
observed for the residual blue side absorption.

We also consider a second model in which the H I and MgII absorption
are used to determine the turbulent and thermal properties of the gas,
and it is the SiIII which resides in a second H I component.  This
model predicts a temperature of T$= 2.02 \times 10^{4}$ K and a
turbulent velocity of 11.4 \kms.
In this scenario, the predicted deuterium velocity width is $b=17.2
\pm 0.5$ \kms, which is 1.9 $\sigma$ lower than that observed for the
extra blue side absorption.

We do not know which of the above scenarios is more likely, as we do
not have a single ionization model which can account for all of the
observed columns and b values.  We can say with certainty, however,
that if the extra absorption on the blue side of the LLS is D I, it
must have a $b < 22.5$ \kms, which is the $b$ value of the main H I
absorption.  It should be noted that in each of the previous five
detections of D/H in a QSO absorption line system the deuterium width
was significantly narrower than the hydrogen width because the H and D
line widths were found to be consistent with almost pure thermal
broadening.  We would thus be quite surprised if the D absorption in
this system has $b_D > 19.2 \pm 0.8$ \kms -- the maximum value
plausible if the D/H absorption arises in the same gas as any of the
observed metal lines.

We also note that because the MgII and SiIII do not have the same widths,
there is probably more than one ionization state to this system.  As a
result, the ionization calculations done in \citet{tytler99} may give an
accurate metalicity for the system.

\section{Is the residual blue side absorption deuterium?}

Given that the new STIS longslit data has specified the absorption
parameters of the neutral hydrogen, we are now in a position to make a
more critical assessment of the identity of the residual blue side
absorption.

We define two simple requirements that the absorption must satisfy if
it is deuterium.  First, the gas must appear at the deuterium position
relative to hydrogen in velocity space, $v = -82$ \kms .  Second, the
velocity width of the gas must be less than the hydrogen, i.e.  it
must have $b < 22.5$ \kms, and we expect $b < 19.2$ \kms.  Our best
fit parameters for the blue side absorption give a position relative
to hydrogen of $v = -75.1 \pm 8.8$ \kms\ and a line width of $b = 27.1
\pm 5.2$ \kms.

We wish to ascertain the likelihood that a D line with physically
plausible parameters would cause us to measure the line parameters
noted above.  To do this, we step through a grid of velocity widths
and relative positions for the extra absorption on the blue side of
\lya.  At each grid point we minimize the $\chi^2$ of the model by
varying the column density of the blue side absorber.  In this way, we
produce contours showing the probability that the data can be
explained by a line at a given velocity with a given b-value.  This
contour plot is shown in Figure 7, with the contours at the 68.3\%,
95.4\%, 99.73\% and 99.99\% confidence levels.

From Figure 7, we again note that the most likely solution is
marginally consistent with our two requirements for the absorption to
be deuterium.  All possible solutions consistent with our deuterium
requirements are outside the 74\% confidence interval from the most
likely solution in the 2-D velocity/velocity dispersion space of
Figure 7.  We note that unless the velocity width of the deuterium is
identical to the velocity width of the hydrogen the probability that
the blue side absorption is not D is $> 74$\%.  We also note that in
the five D/H systems previously measured, the deuterium has always
been significantly narrower than the hydrogen.

If we require the deuterium be consistent with the thermal parameters
estimated from the SiIII and H I lines, i.e. $b_{D} = 19.2 \pm 0.8$
\kms, the blue side absorption is not D at a 98\% confidence level.
Because no system has ever been observed with purely turbulent line
widths, and because both the SiIII and MgII have significantly smaller
line widths than the H I, we feel that $b_{D} = 19.2$ \kms\ is the
highest likely $b$ value for a deuterium line associated with this
system.  We thus conclude that the extra absorption is not D with a
confidence $> 98$\%.

If instead the velocity width of the deuterium is consistent with the
photoionization calculations and MgII+H I velocity widths, i.e. $b_{D}
= 17.2 \pm 0.5$ \kms , the deuterium only solution is less than
$10^{-4}$ times as likely than the best solution describing the data.

We also investigated two component Hydrogen models for this system,
under the assumption that the discrepant MgII and SiIII $b$ values may
indicate a more than one major component.  We found that the total
hydrogen absorption profile is well constrained by the high order
Lyman series lines, and as a result introducing a multi-component
hydrogen model does not increase the likelihood that the extra blue
side absorption is deuterium.

\section{Discussion}

This system has received extensive consideration because it was
thought to be possible that it would provide evidence for a very high
D/H ratio, in contrast to lower values of D/H found in all other QSO
absorption systems.  The absorption system was believed to be a single
component system, which would have indicated that the extra blue side
absorption was D.

However, now that we have properly constrained the main hydrogen
absorption through profile fitting of it's high order Lyman series
lines, it now appears likely that there is contaminating absorption on
the blue side of the LLS.  This conclusion is supported by the fact
that no physically plausible D line can explain the data and that
there is also extra absorption on the red side of the system.  Since
the red absorption can not be deuterium, it indicates the LLS has the
type of complex velocity structure that can easily produce extra
absorption outside of the main component.

We thus conclude that this is a multi component absorption system
which likely has unconstrained absorption near the expected position
of deuterium, and that no determination of the primordial D/H ratio
can be obtained from \object\ with data available at this time.

\section{Acknowledgments}

We are grateful to Kailash Sahu for expert advice on the STIS line
spread function.  This work was funded in part by STScI grant GO-7292
and by NSF grant AST-9900842.  The MgII spectra were obtained from the
W.M. Keck observatory, which is managed by a partnership among the
University of California, Caltech and NASA.  We are grateful to Steve
Vogt, the PI for the Keck HIRES instrument, and to the W.M. Keck
Observatory staff.

\clearpage

\begin{figure}
\figurenum{1}
\plotone{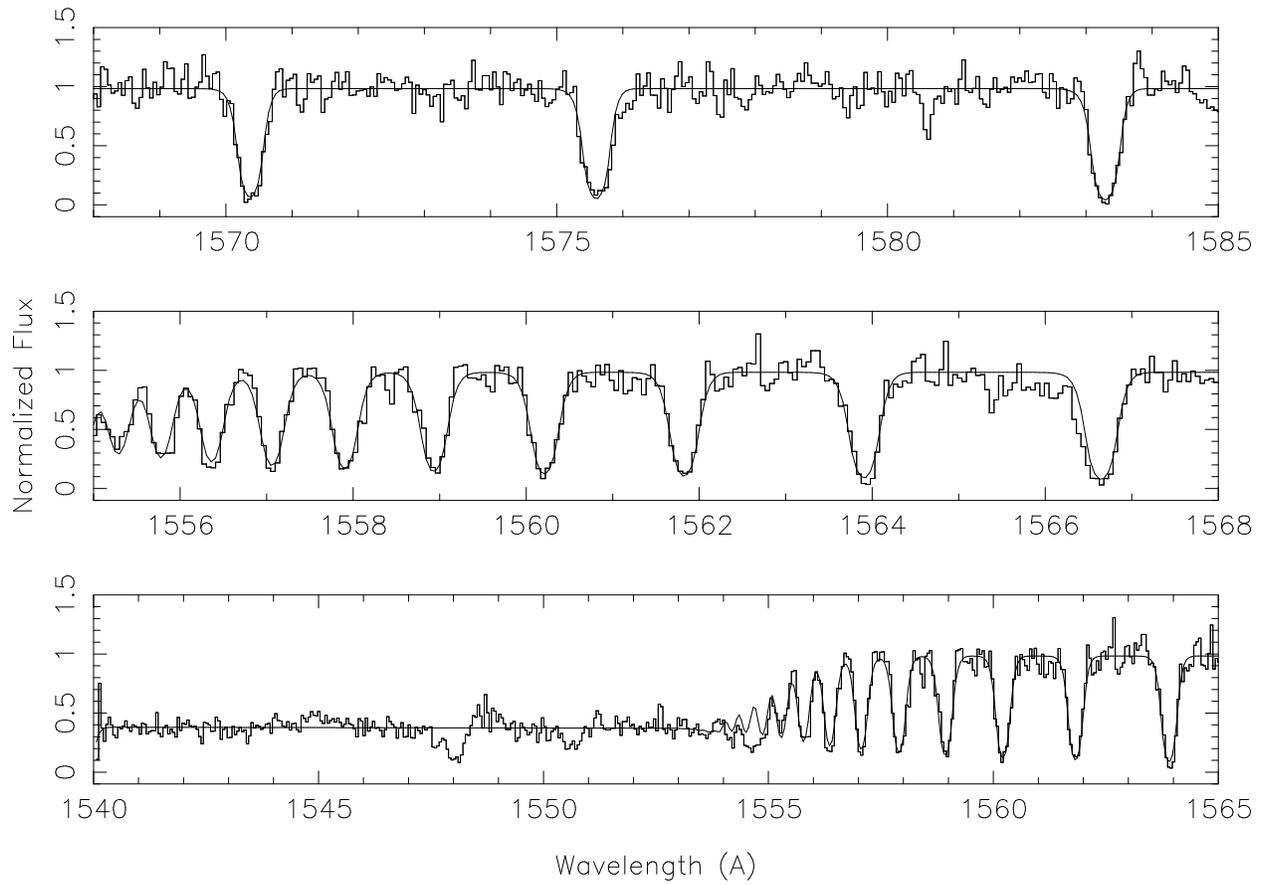}
\caption{STIS G140M spectrum of the Lyman limit region of \obja.  The
smooth line is our best fit solution to the hydrogen absorption.}
\end{figure}

\begin{figure}
\figurenum{2}
\plotone{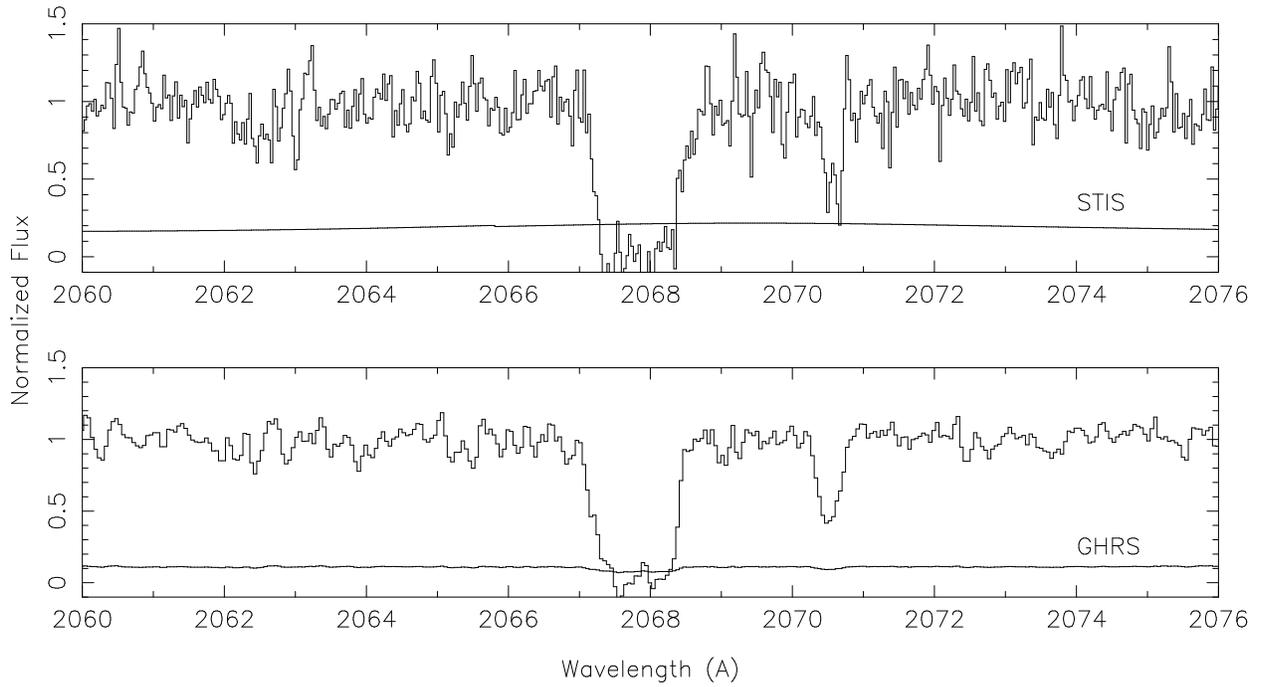}
\caption{The absorption at $\sim 2068$ \AA is \lya\ at \zabs\ = 0.7011
seen in the STIS E140M (top, 9.3 \kms\ FWHM) and GHRS (bottom, 14.1 \kms\ FWHM)
spectra of \obja.  A long stretch of spectra is shown 
to demonstrate the quality of the data.  The line near 0.1
indicates the error on the normalized flux in each pixel in the spectrum.}
\end{figure}

\begin{figure}
\figurenum{3}
\plottwo{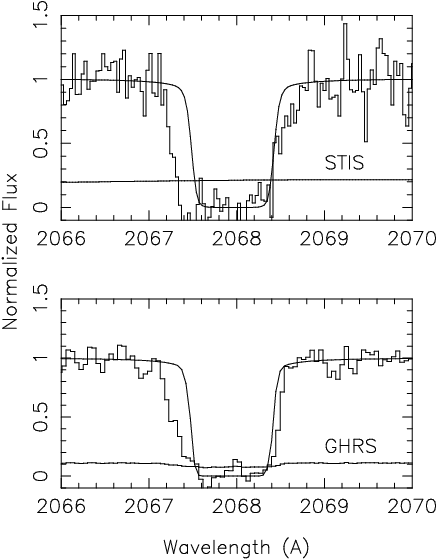}{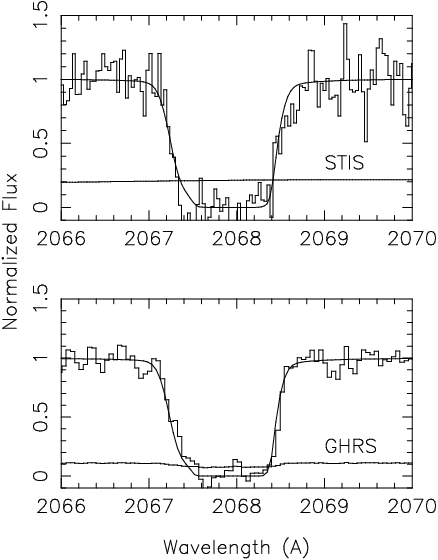}
\caption{The \lya\ absorption line in the STIS and GHRS spectra.  The
left hand panels show the expected H I absorption from the fit to the
Lyman limit.  The right hand panels show our full model to the data,
summarized in Table 2.}
\end{figure}

\begin{figure}
\figurenum{4}
\epsscale{0.6}
\plotone{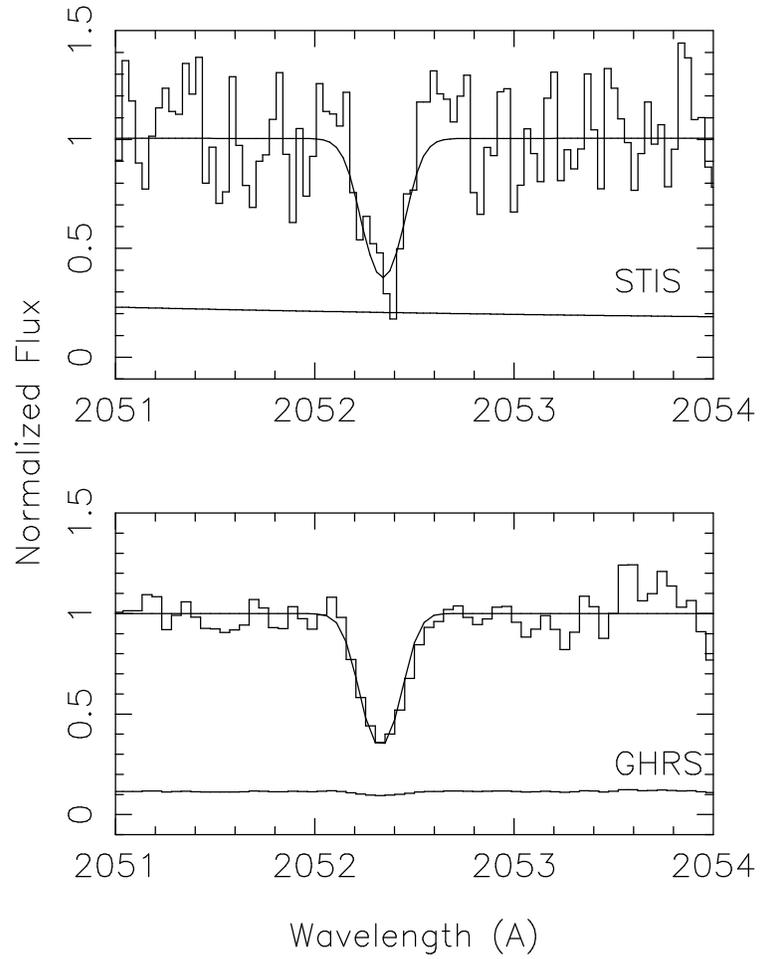}
\epsscale{1.0}
\caption{The observed SiIII (1206) absorption in the STIS and GHRS
spectra, along with the single best fit to both.}
\end{figure}

\begin{figure}
\figurenum{5}
\epsscale{0.6}
\plotone{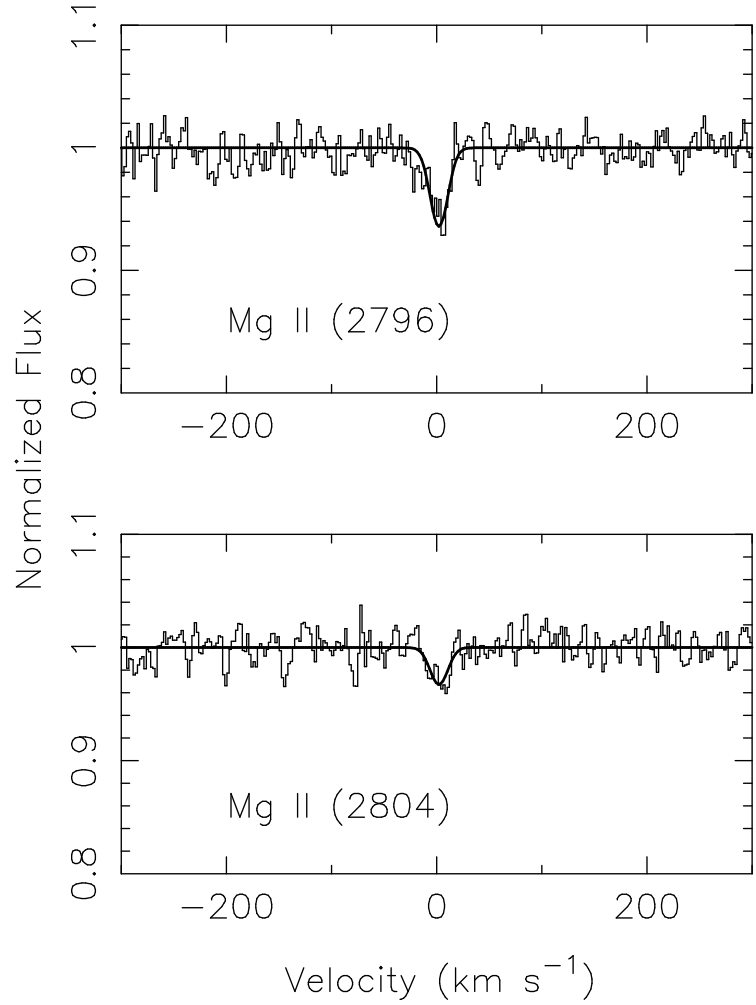}
\epsscale{1.0}
\caption{The MgII absorption in the Keck HIRES spectra.  The velocity zero
point is the redshift of the hydrogen LLS, \zabs =
0.701084.}
\end{figure}

\begin{figure}
\figurenum{6}
\epsscale{0.7}
\plotone{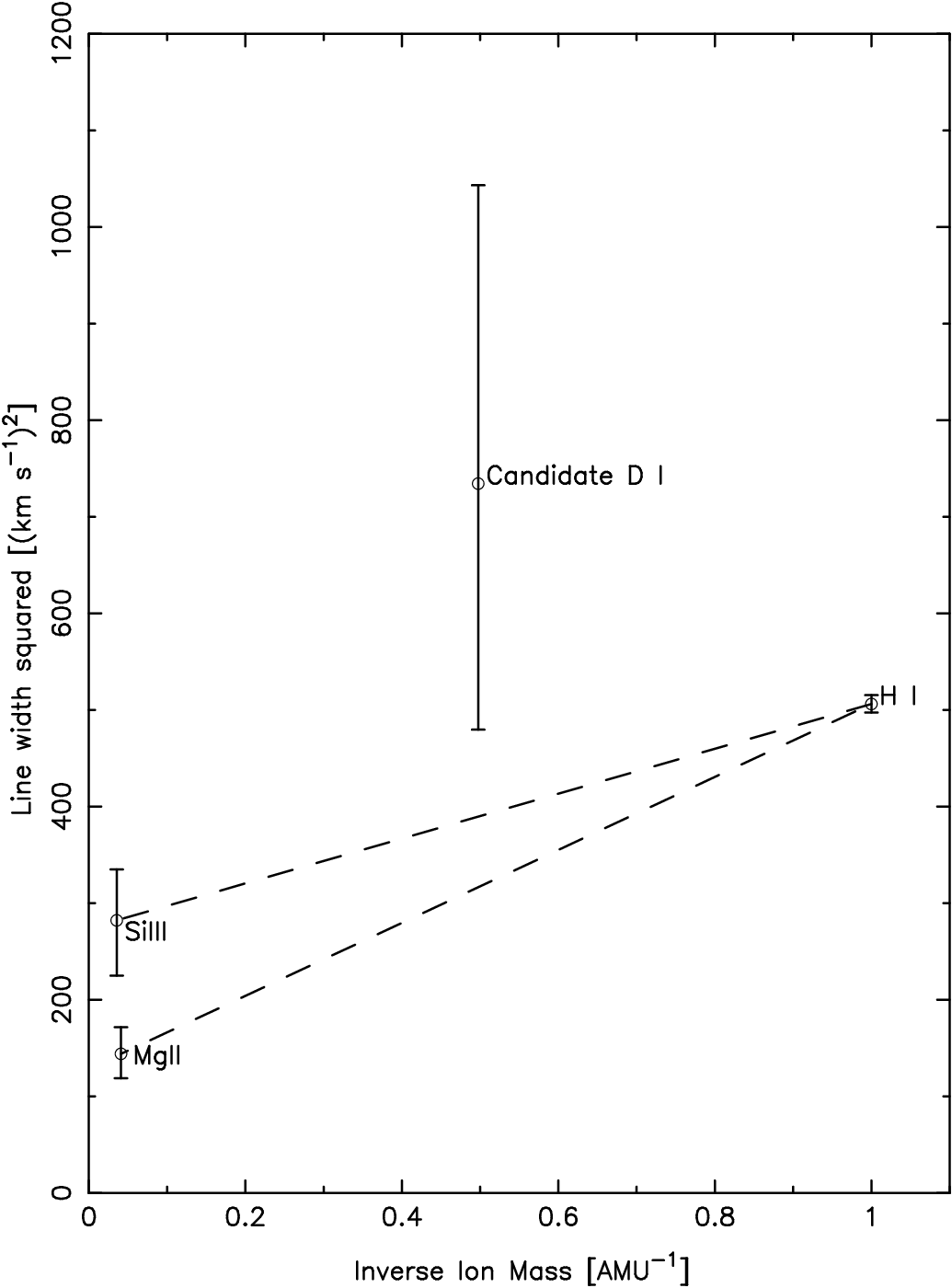}
\epsscale{1.0}
\caption{The line widths of the observed ions.  If all of the
absorption came from the same gas, all of the points should lie on a
strait line.  The fact that H I, SiII and MgII do not lie on a strait
line indicates that the system contains more than one thermal state
(or measurement errors).  If either the MgII or the SiIII comes from
the gas showing H I, then the residual blue side absorption (here
marked `candidate D I') is wider than expected for D.}
\end{figure}

\begin{figure}
\figurenum{7}
\plotone{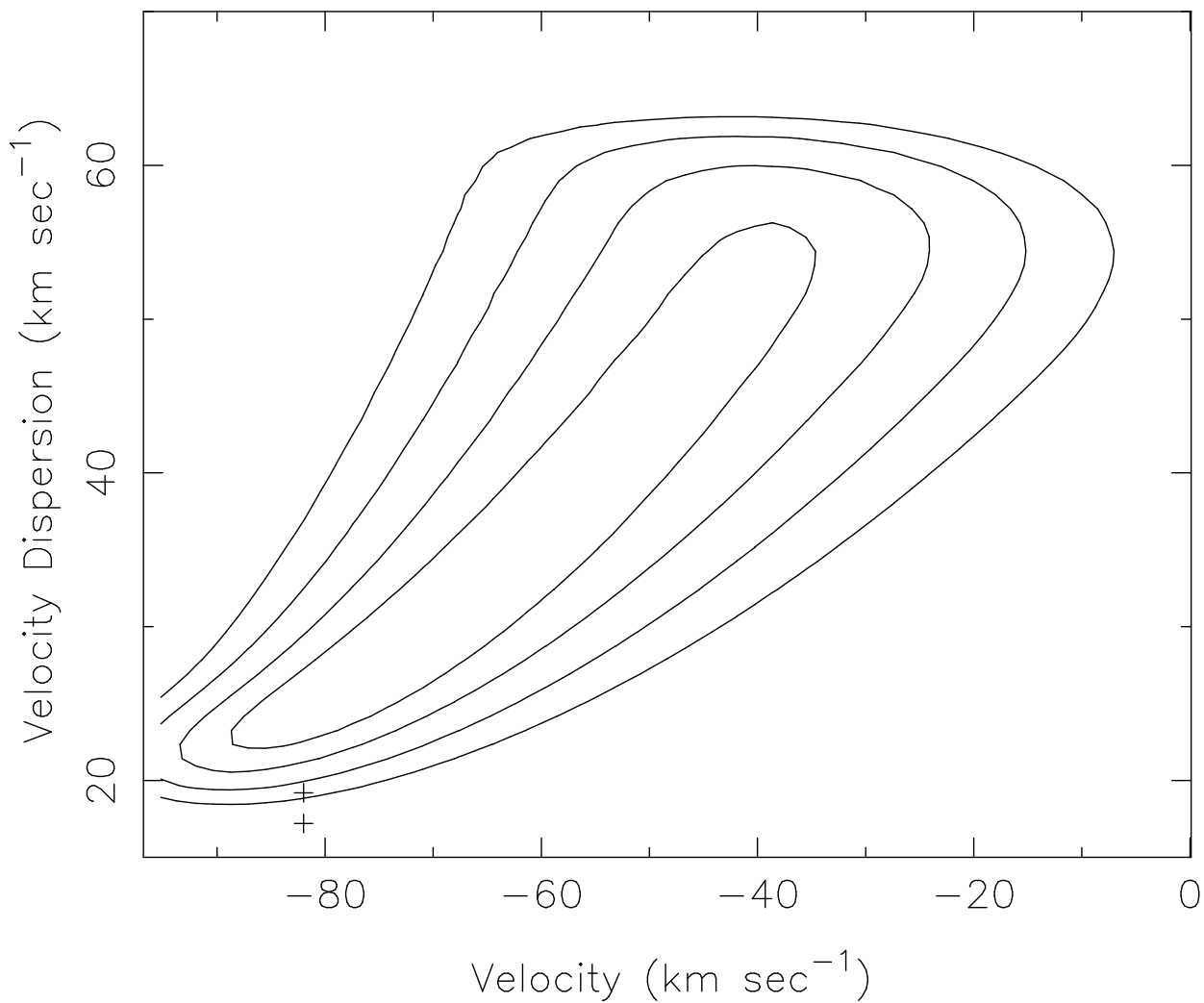}
\caption{Probability that the blue-side absorption has a given
velocity dispersion and redshift.  The contour levels show the 68.3\%,
95.4\%, 99.73\% and 99.99\% confidence levels based on differences in
$\chi^2$ of 2.3, 6.17, 11.8 and 18.4 from the best fit solution.  The
two marked points mark the location in parameter space of possible D
lines -- the point at $b = 17.2$ \kms\ is the D line expected if the H
I and MgII absorption arise in the same gas, and the $b = 19.2$ \kms\
point is the D line expected if the H I and SiIII absorption arise in
the same gas.}
\end{figure}

\clearpage

\begin{deluxetable}{lll}
\tablecaption{STIS normalized line spread functions}
\tablewidth{0pt}
\tablehead{
\colhead{Pixel} &
\colhead{e230} &
\colhead{g140m}
}

\startdata
   0   &     0.3765     &     0.3059 \\
   1   &     0.2201     &     0.1702 \\
   2   &     0.0589     &     0.0849 \\
   3   &     0.0204     &     0.0564 \\
   4   &     0.0080     &     0.0192 \\
   5   &     0.0030     &     0.0093 \\
   6   &     0.0013     &     0.0048 \\
   7   &                &     0.0022

\enddata
\end{deluxetable}

\begin{deluxetable}{cccccccccc}
\tablecaption{Lines observed in the \zabs $\sim$ 0.7011 LLS}
\tablewidth{0pt}
\tablehead{
\colhead{Ion} &
\colhead{log(N)} &
\colhead{$b$} &
\colhead{$z$} &
\colhead{velocity} &
\colhead{$\sigma_{\rm log(N)}$} &
\colhead{$\sigma_b$} &
\colhead{$\sigma_z$} &
\colhead{$\sigma_v$} \\
& \colhead{(\cmm)} & \colhead{(\kms)} & & \colhead{(\kms)}
}

\startdata
H I     & 17.23  & 22.5  & 0.701084  &     0  & 0.005  &  0.2   & 2e-6  & 0.35  \\
H I     & 13.47  & 22.2  & 0.701406  &  56.7  & 0.8    & 13.1   & 1e-4  & 16.1  \\
H I\tablenotemark{a} & 13.84  & 27.1  & 0.700658  & -75.1  & 0.14   &  5.2   & 5e-5  &  8.8  \\
SiIII   & 12.85  & 16.8  & 0.701062  &  -2.1  & 0.04   &  1.8   & 8e-6  &  1.4  \\
MgII    & 11.5   &  12.0  & 0.701100  &   2.8  & 0.05    &  1.1   & 5e-6  &  0.88 \\
SiIV\tablenotemark{b} & $<12.6$ & -- & -- & -- & 0.1 & -- & -- & --  
\enddata
\tablenotetext{a}{Possibly Deuterium}
\tablenotetext{b}{Not detected}
\end{deluxetable}

\end{document}